\documentstyle[11pt,newpasp,twoside,epsf]{article}
\begin{document}
\title{ Absolute Lineshifts \\
-- a new diagnostic for stellar hydrodynamics}
\author{ Dainis Dravins }
\affil{ Lund Observatory, Box 43, SE-22100 LUND, Sweden }

\begin{abstract} For hydrodynamic model atmospheres, {\it absolute lineshifts} are becoming an observable diagnostic tool beyond the classical ones of line-strength, -width, -shape, and -asymmetry.  This is the wavelength displacement of different types of spectral lines away from the positions naively expected from the Doppler shift caused by stellar radial motion.  Caused mainly by correlated velocity and brightness patterns in granular convection, such absolute lineshifts could in the past be studied only for the Sun (since the relative Sun-Earth motion, and the ensuing Doppler shift is known).  For other stars, this is now becoming possible thanks to three separate developments: (a) Astrometric determination of stellar radial motion; (b) High-resolution spectrometers with accurate wavelength calibration, and (c) Accurate laboratory wavelengths for several atomic species.  Absolute lineshifts offer a tool to segregate various 2- and 3-dimensional models, and to identify non-LTE effects in line formation.
\end{abstract}


\section{Progress in understanding stellar atmospheres}

Improved understanding of physical processes in stellar atmospheres comes from the interplay between theory and observations.  The validity of, e.g., hydrodynamic simulations of stellar photospheres can be tested by whether they succeed in reproducing observed shapes and asymmetries of photospheric spectral lines.  On one hand, the success of such models (at least for solar-type stars) is encouraging in suggesting that many physical processes probably are beginning to get understood.  However, on the other hand, different models with dissimilar parameter combinations sometimes are able to produce similar results, showing the need for further observational constraints.  New diagnostic tools are needed to, e.g., segregate 2-dimensional models from 3-dimensional ones, to identify effects of non-LTE, or of 3-dimensional radiative transfer in line formation.

One has to remember that progress in astronomy (and natural science in general) depends upon the {\it falsification} of theoretical concepts and hypotheses.  If the ambition is to improve a model, the aim is {\it not} to have it fit the observations, but rather to identify situations where the model and observations are in {\it disagreement} (in the case of agreement, there is really nothing more that can be learned).  A route for progress in stellar-atmospheres research is then to identify new diagnostic tools that both can be accurately observed and be precisely predicted by models.  A confrontation between these observational and theoretical measures may then segregate among competing concepts and guide models toward greater realism and sophistication.

This paper discusses one such new diagnostic tool: 'Absolute' wavelength shifts of photospheric spectral lines.  This denotes the wavelength displacement of spectral lines away from the positions they naively would be expected to have, had the lines been affected by only the Doppler shift caused by the relative star-Earth motion.  In the past, such absolute lineshifts could be studied only for the Sun (since the relative Sun-Earth motion, and the ensuing Doppler shift is known from planetary-system dynamics, and does not rely on measurements in the solar spectrum).  For other stars, this is now becoming possible thanks to three separate developments: (a) Accuracies reached in space astrometry has made it possible to accurately determine stellar radial motion from second-order effects in astrometry only, without using any spectroscopy; (b) The availability of high-resolution spectrometers with accurate wavelength calibration (such as designed for exoplanet searches), and (c) Accurate laboratory wavelengths for several atomic species.  Already the two latter points (b) and (c) permit studies of {\it differential} shifts between different classes of lines in one and the same star even if, in the absence of astrometric radial velocities, the shifts can not be placed on an absolute scale.

Asymmetries and wavelength shifts of photospheric absorption lines originate from correlated velocity and brightness patterns (stellar granulation): rising (blueshifted) elements are hot (bright), and convective blueshifts normally result from a larger contribution of such blueshifted photons than of redshifted ones from the sinking and cooler (darker) gas.  The exact amount of shift differs among lines with different conditions of formation.  For example, high-excitation lines may form predominantly in the hottest (also the most rapidly rising, and most blueshifted) elements, and thus show a more pronounced blueshift.  By contrast, lines formed in high-lying layers of convective overshoot may experience an inverted correlation, instead resulting in a convective redshift.  For overviews of processes causing such shifts, see Dravins (1999), Asplund et al.\ (2000), Allende Prieto et al.\ (2002), and references therein.  One observational advantage of lineshifts (as opposed to line asymmetries or precise line profile shapes) is that they may be simpler to measure also at modest spectral resolutions, when the bisectors describing line asymmetries may already be strongly perturbed (Dravins \& Nordlund 1990b; Hamilton \& Lester 1999).

\section{Radial velocities without spectroscopy!}

Traditionally, stellar radial velocities have been determined by spectroscopy, utilizing the Doppler effect.  The advent of high-accuracy (milli-arcsecond) astrometric measurements now permits radial velocities to be obtained also by purely geometric methods.  Such {\it astrometric radial velocities}, giving the motion of the stellar center-of-mass, are independent of phenomena affecting stellar spectra, such as stellar pulsation, convection, rotation, winds, isotopic composition, pressure, and gravitational potential.  Among the several astrometric techniques (Dravins et al.\ 1999b), one method reaches sub-km~s$^{-1}$ accuracies already with existing data.  This 'moving-cluster method' is based on the circumstance that stars in suitable open clusters move through space with a common [average] velocity vector.  The radial-velocity component makes the cluster appear to contract or expand due to its changing distance.  This relative rate of apparent contraction (observed through stellar proper motions) equals the relative rate of change in distance, which can be converted to a linear velocity (in km~s$^{-1}$) if the distance is known from trigonometric parallaxes.  Once the space velocity is known, the radial velocity for any member star follows by projecting the cluster's velocity vector onto the line of sight.  The method can be regarded as an inversion of the classical moving-cluster problem, where the distance is derived from proper motions and (spectroscopic) radial velocities: here, by instead first knowing the distances, the radial velocities follow: Figure 1.

\begin{figure}
\plotfiddle{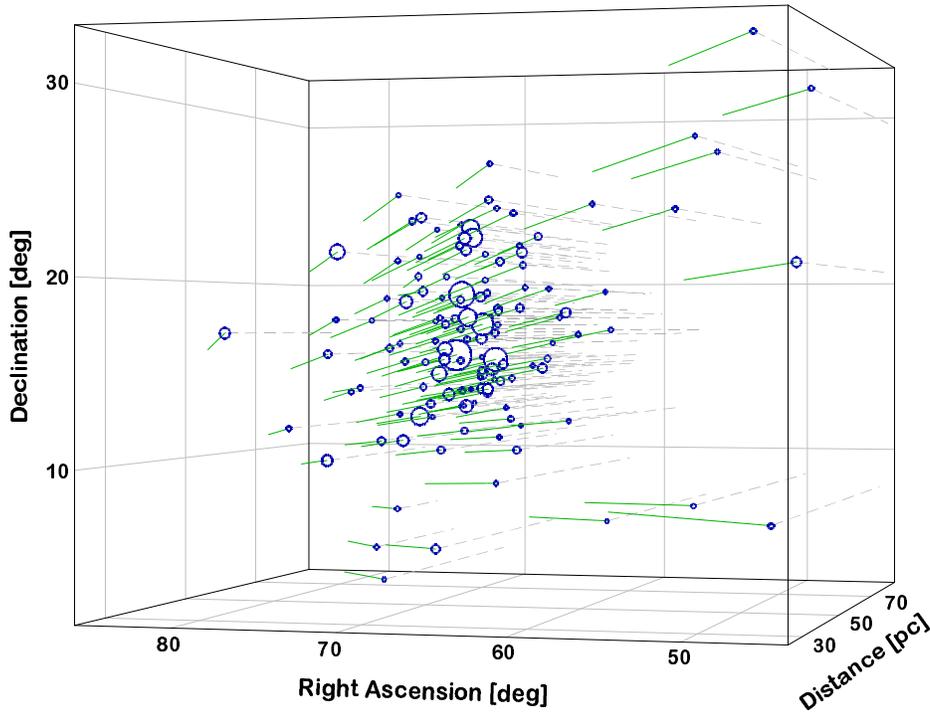}{9.8cm}{0}{55}{55}{-255}{-145}
\caption{Proper motions in the Hyades as measured by Hipparcos, with symbol size indicating stellar brightness.  Stars in a moving cluster share the same average velocity vector.  Parallaxes give the distance, while proper-motion vectors show the fractional change with time of the cluster's angular size.  The latter corresponds to the time derivative of distance, yielding the radial velocity (Dravins et al.\ 1997)}
\end{figure}

\section{Lineshifts intrinsic to stellar atmospheres}

On Hipparcos, an observing program was carried out for stars in moving clusters, yielding astrometric radial-velocity solutions for more than 1,000 stars (Madsen et al.\ 2002), including about one hundred stars in the Hyades and Ursa Major groups.  The error budgets are somewhat complex, but typical accuracies reach around 0.5 km~s$^{-1}$.

Differences between astrometrically determined radial velocities (giving true motions of the stellar centers of mass) and the apparent spectroscopic velocities of different features reveal lineshifts intrinsic to stellar atmospheres, such as convective and gravitational lineshifts.  Figure 2 for Hyades stars indicates that F-type spectra have enhanced convective blueshifts relative to cooler ones (as expected from hydrodynamic simulations; Dravins \& Nordlund 1990ab; Allende Prieto et al. 2002), a tantalizing trend that seems to continue to even hotter stars (where corresponding hydrodynamic models are not yet developed).

\begin{figure}
\plotfiddle{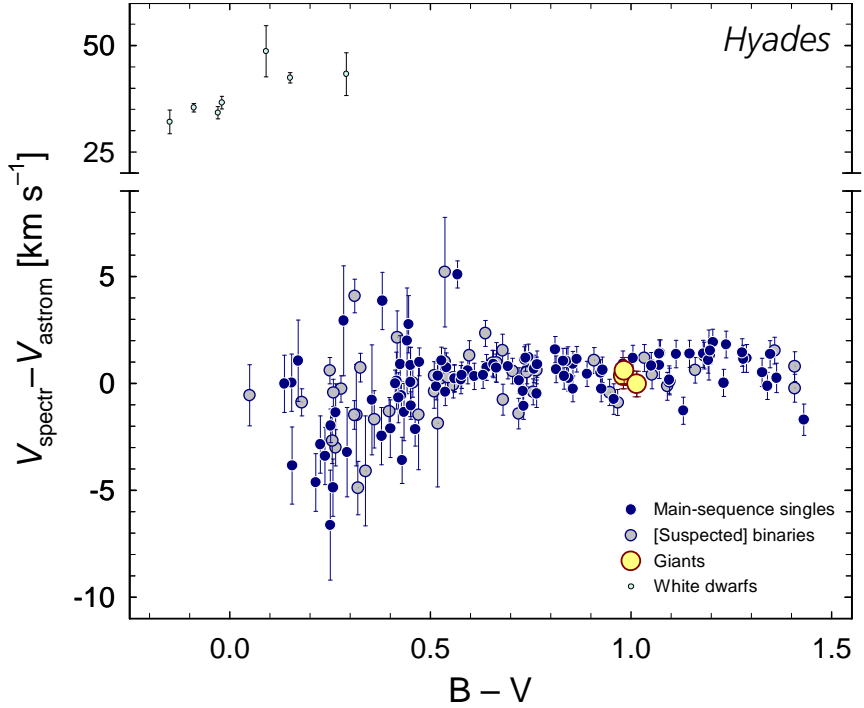}{9.2 cm}{0}{60}{60}{-185}{-225}
\caption{ The Hyades: Differences between spectroscopic radial-velocity values from the literature, and astrometric determinations.  An increased blueshift of spectral lines in stars somewhat hotter than the Sun ($\bv \simeq 0.3-0.5$) is theoretically expected due to their more vigorous surface convection, causing greater convective blueshifts.  Gravitational redshifts of white-dwarf spectra place them far off main-sequence stars.  The error bars show the combined spectroscopic and astrometric errors (Madsen et al.\ 2002)}
\end{figure}

\section{Gravitational redshifts}

Another phenomenon affecting the difference between spectroscopic and astrometric radial velocities is the gravitational redshift, $V_{\rm grav} =GM/Rc$, involving the constant of gravitation, the speed of light, and the stellar mass and radius.  Using accepted solar values, we obtain 636 m~s$^{-1}$ for light escaping from the solar photosphere to infinity, and 633 m~s$^{-1}$ for light intercepted at the Earth.  A spectral line formed at chromospheric heights (30 Mm, say), will have this shift decreased by some 20 m~s$^{-1}$, and a coronal line by perhaps 100 m~s$^{-1}$.  For other stars, the shift scales as ($M/M_{\odot }$)($R_{\odot }/R$).

\begin{figure}
\plotfiddle{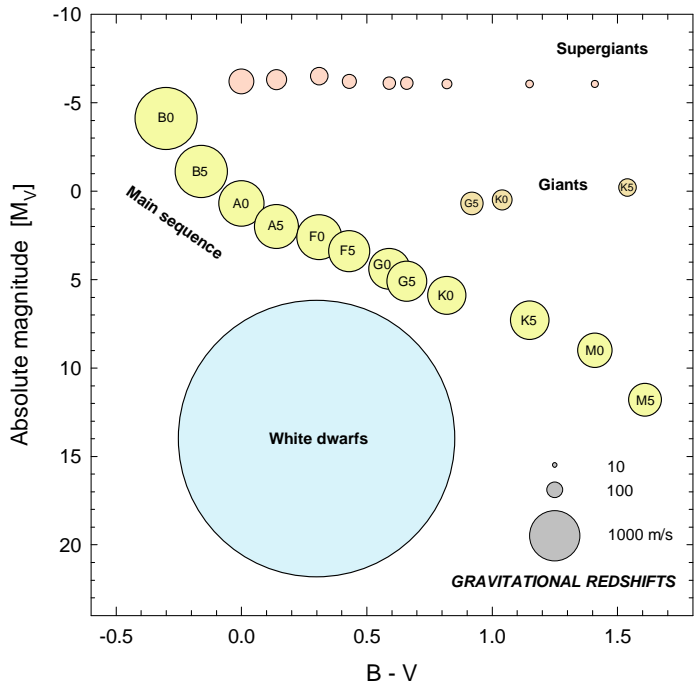}{9.7 cm}{0}{50}{50}{-140}{-110}
\caption{ {\it Gravitational redshifts} throughout the Hertzsprung-Russell diagram change by three orders of magnitude between white dwarfs ($\simeq$ 30 km~s$^{-1}$) and supergiants ($\simeq$ 30 m~s$^{-1}$).  Symbol area is proportional to the amount of shift (Dravins et al.\ 1999a)}
\end{figure}

It is believed that this shift can be reasonably well predicted from stellar structure models (for well-studied single main-sequence stars to an accuracy of perhaps 50 m~s$^{-1}$), and it is not expected to change significantly between different photospheric lines.  Therefore, it should be much less of an unknown variable than the convective lineshifts.

Figure 3 shows expected gravitational redshifts predicted from standard models.  The shift does not vary much on the main sequence between A5 V and K0 V ($\simeq$~650 m~s$^{-1}$), but reaches twice that value for more massive early-B stars with $M \simeq$~10$~M_{\odot }$.  However, the convective lineshift, $V_{\rm conv}$, is expected to be about $-$1000 m~s$^{-1}$ for F5~V, $-$400 m~s$^{-1}$ for the Sun, and $-$200 m~s$^{-1}$ for K0~V (Dravins 1999, and references therein).  The difference $V_{\rm spec}-V_{\rm astrom}$ can thus be expected to lie in the range between $-$400 and +400 m~s$^{-1}$, when moving along the main sequence from F5~V to K0~V (\bv ~between 0.4 and 0.9).

\section{3-D models of convective shifts in different stars}

Synthetic line profiles from three-dimensional hydrodynamical models do well reproduce photospheric line shapes and shifts in solar-type stars.  Figure 4 serves to (a) illustrate the magnitude of the effect, and (b) to show that the intensity profile alone (at least on first sight) does not reveal any striking signature of the complex convective velocity fields shaping it.

\begin{figure}
\plotfiddle{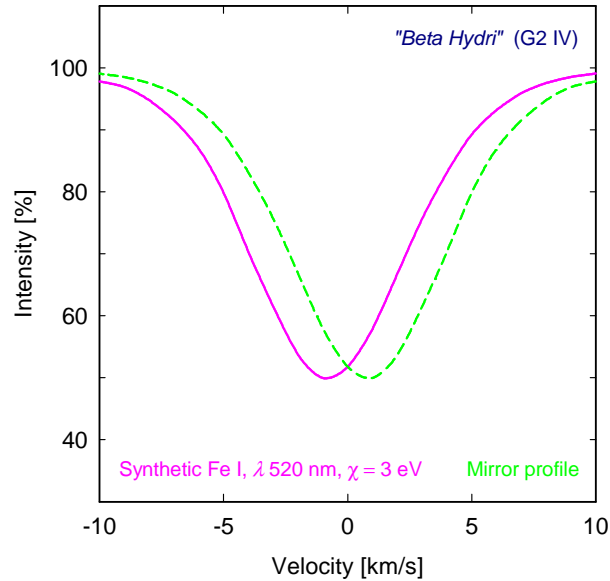}{8.5 cm}{0}{55}{55}{-180}{-190}
\caption {A model with solar temperature but one quarter of its surface gravity (corresponding to the subgiant $\beta$ Hyi), has smaller photospheric pressure but greater granular velocities (by a factor of 1.5-2), since the same energy flux as in the Sun must be carried by lower-density gas.  Its resulting convective blueshift is greater than in the Sun, here illustrated by a synthetic \ion{Fe}{I} $\lambda$ 520 nm line, mirrored about the zero-point wavelength (adapted from Dravins et al.\ 1993)}
\end{figure}

Figure 5 shows normalized bisectors of synthetic spectral lines in different solar-type stars, obtained from 3-D hydrodynamic models.  It shows how the {\it asymmetry} of such lines may be quite similar, although their wavelength shifts differ by a full factor of 5!  Clearly, lineshifts are sensitive signatures of dissimilar convection patterns in different stars.  In particular, the more vigorous convection in subgiants produces a stronger convective blueshift than in dwarfs, despite quite similar line profile shapes (Dravins \& Nordlund 1990a).

\begin{figure}
\plotfiddle{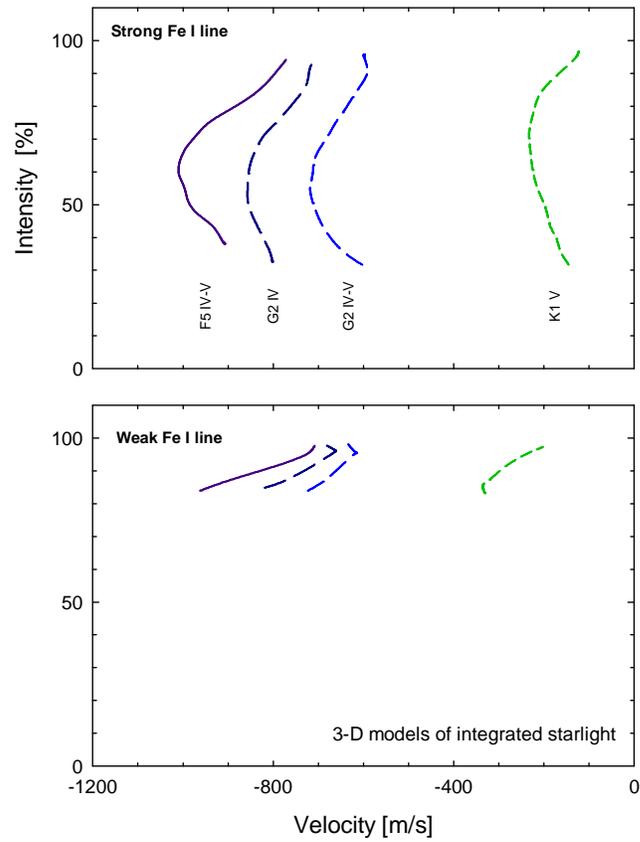}{12 cm}{0}{60}{60}{-180}{-110}
\caption {Asymmetries of the same spectral line in different stars: \ion{Fe}{I} bisectors from four different 3-D simulations representing Procyon (F5 IV-V), $\beta$~Hyi (G2 IV), $\alpha$~Cen~A (G2 IV-V), and $\alpha$~Cen~B (K1~V).  Convective blueshift increases with increasing temperature, and increasing luminosity (adapted from Dravins \& Nordlund 1990b)}
\end{figure}

\begin{figure}
\plotfiddle{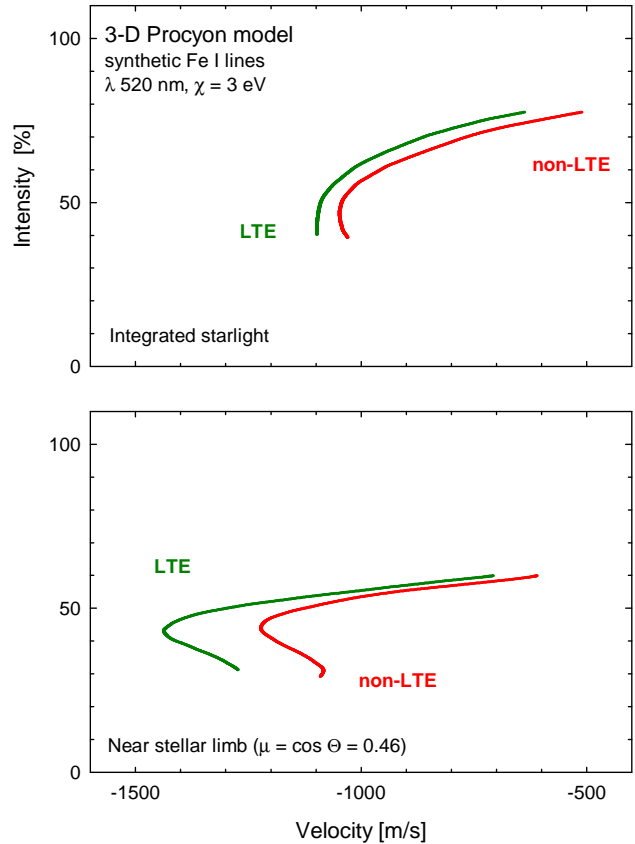}{12cm}{0}{60}{60}{-165}{-125}
\caption {Non-LTE effects in convective lineshifts of \ion{Fe}{I} lines in a 3-D simulation of Procyon.  Although line shapes and asymmetries are quite similar, there are differences in their shifts.  The smaller shift in non-LTE is caused by a selective reduction by the most blueshifted line components above the hottest granules, whose intense ultraviolet flux ionizes the gas above, thus decreasing contributions from neutral species (adapted from material in Dravins \& Nordlund 1990a)}
\end{figure}

\section{Non-LTE diagnostics}

Lineshifts offer a non-LTE diagnostic for synthetic line profiles, since such effects may show up more clearly in shifts than in either profiles or asymmetries.  Although line {\it shapes} may well be similar in LTE, line {\it shifts} can be different.  For lines in the visual from neutral metals in solar-type stars, non-LTE calculations normally reveal a smaller convective blueshift because the fastest upflows occur in the hottest granular elements.  These emit strong ultraviolet flux, ionizing the gas above, thus decreasing the most blueshifted line absorption contributions from neutral species.  Close to the stellar limb, the bisector effects become more significant, an effect of a 'corrugated' stellar surface (Figure 6).

\section{ Solar lineshift patterns}

For the Sun, absolute lineshift studies have been possible since long ago.  Of course, models for the solar atmosphere are developed to a much higher sophistication than those for other stars, implying that more detailed model predictions can be made and tested.  Observational limits are set by systematic errors in wavelength scales which, even in the best solar spectrum atlases, correspond to about 100 m~s$^{-1}$ or worse; comparable to the noise in the best sets of laboratory wavelengths for low-excitation lines from neutral metals.  The most sensitive tests of various models might be with high-excitation lines, lines from ionized species, or from molecules.  Unfortunately, such lines also appear to be those most sensitive to systematic errors in the determination of laboratory wavelengths.  Nevertheless, improved laboratory measurements make an increasing number of lines available for study.  Figure 7 shows the pattern of convective lineshifts for \ion{Fe}{II} lines at solar disk center, using recent laboratory data.

\begin{figure}
\plotfiddle{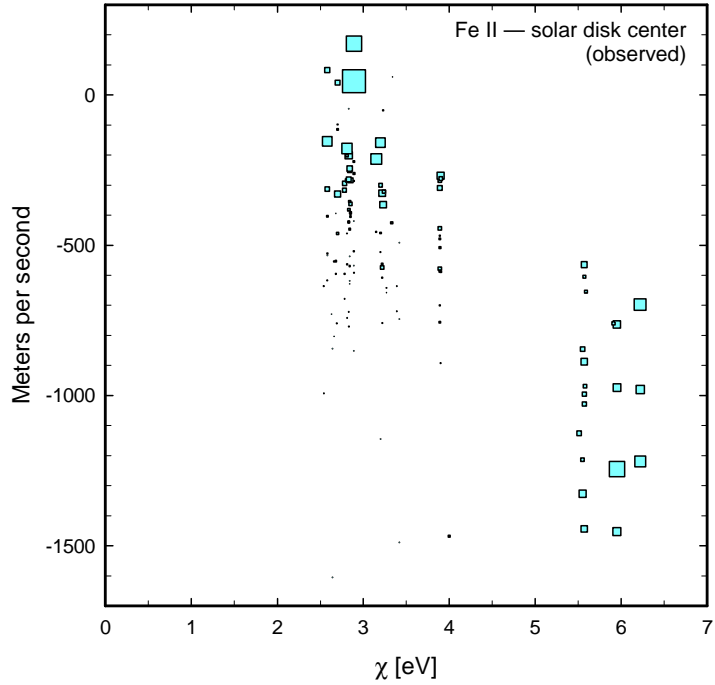}{9.5cm}{0}{50}{50}{-160}{-40}
\caption {Convective blueshifts for 137 \ion{Fe}{II} lines in the visual at solar disk center, as function of oscillator strength (symbol size) and lower excitation potential.  A value of '0' means that the solar wavelength equals the laboratory one, corrected for solar gravitational redshift, and for the Doppler shift due to the relative Sun-Earth motion.  Solar wavelengths: Allende Prieto \& Garc{\'\i}a L\'{o}pez 1998; Laboratory data: S.Johansson (Lund Observatory; private comm.)}
\end{figure}

\begin{figure}
\plotfiddle{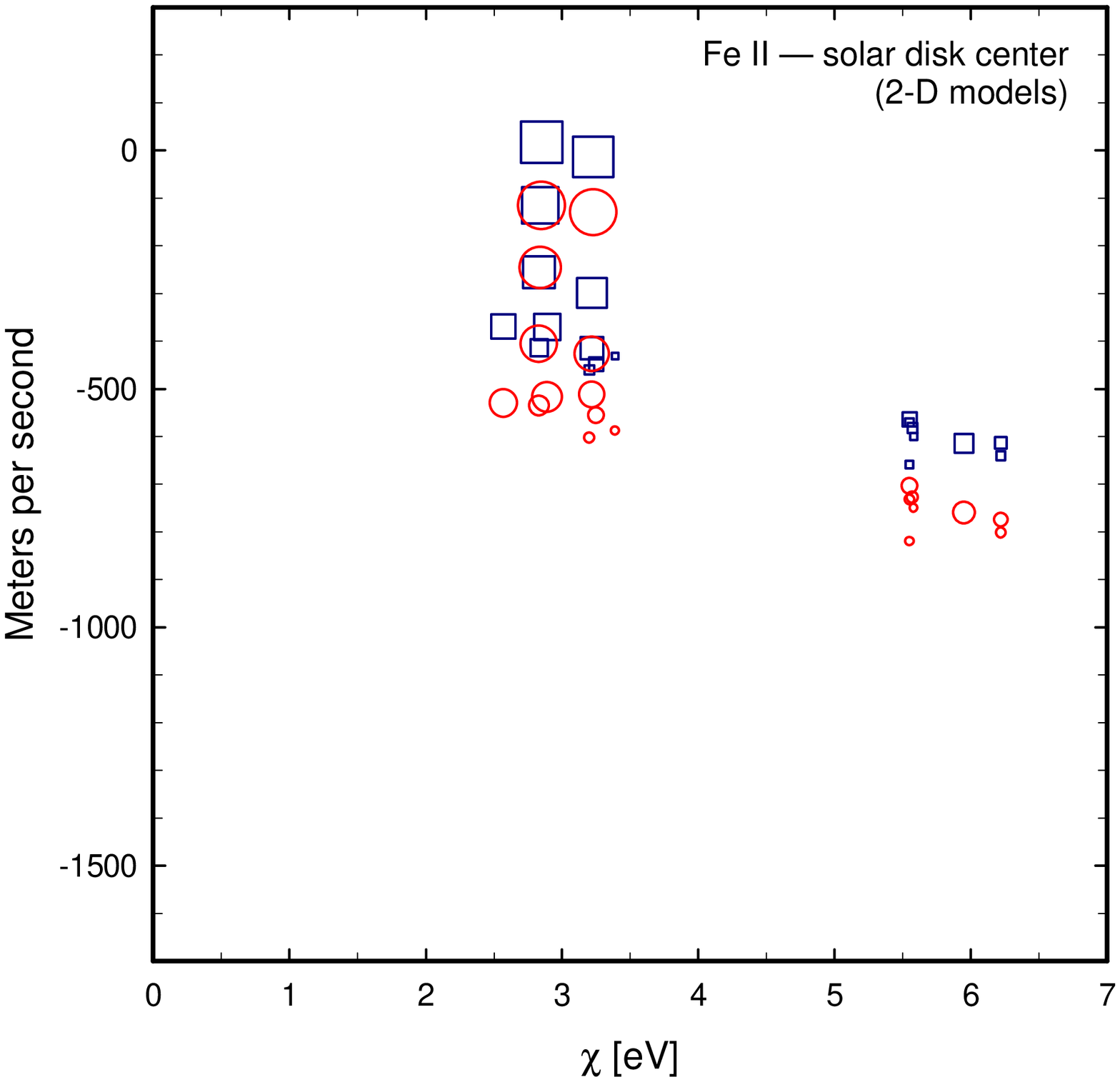}{9.5cm}{0}{50}{50}{-160}{-40}
\caption {Modeled solar \ion{Fe}{II} convective lineshift patterns in the visual.  These line-bottom shifts of 21 specific \ion{Fe}{II} lines at solar disk center were computed from two-dimensional hydrodynamic models.  Squares show results from one particular model, and circles from another one, differing in horizontal extent and depth.  Symbol area is proportional to solar equivalent width.  Calculations by the late Aleksey Gadun (Main Astronomical Observatory, Kyiv, Ukraine; private comm.)}
\end{figure}

Figure 8 is a sample of theoretically predicted convective lineshifts for specific \ion{Fe}{II} lines, a subset of the observed ones in Figure 7.  These line-bottom shifts of 21 \ion{Fe}{II} lines at solar disk center were computed from two-dimensional hydrodynamic models.  Squares show results from one particular model, and circles from another one, differing in horizontal extent and depth.  Although the predicted pattern is qualitatively similar to the observed one in Figure 7, the granulation structure produced by 2-D models is not sufficiently vigorous to fully reproduce the amplitude of observed lineshifts, for which 3-D models thus are required.  (The purpose of this figure is not to argue for any specific degree of agreement between theory and observation, but to demonstrate how different hydrodynamic models may be segregated though the different lineshift patterns predicted.)

The increasing availability of accurate laboratory wavelengths now permits studies for more species than the classically used iron only, including \ion{Ca}{I}, \ion{Co}{I}, \ion{Cr}{I}, \ion{Ti}{I}, and \ion{Ti}{II}.  This both increases the available sample of lines, and probes different conditions of line formation.  As an example, Figure 9 shows the solar lineshift pattern for lines from neutral nickel.

\begin{figure}
\plotfiddle{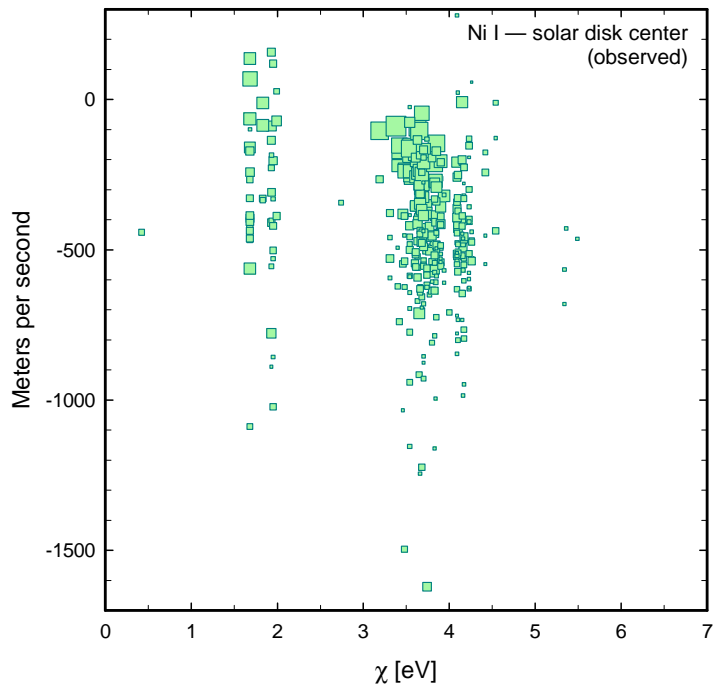}{9.5cm}{0}{50}{50}{-160}{-40}
\caption {Solar lineshift patterns for 369 \ion{Ni}{I} lines in the visual.  Symbol size reflects laboratory intensity.  Solar wavelengths: Allende Prieto \& Garc{\'\i}a L\'{o}pez 1998; Laboratory data: Litz\'{e}n et al.\ 1993}
\end{figure}

\section{ A new diagnostic for stellar atmospheres}

A combination of observational and experimental advances now permit absolute lineshifts to be introduced as a diagnostic tool for stellar atmospheres, beyond the previously established ones of line-strength, -width, -shape, and -asymmetry.

On the theoretical side, lineshift patterns cannot yet be predicted across the whole HR-diagram.  Not only do these require a credible modeling of the atmospheric hydrodynamics, but the lineshifts may vary greatly among different classes of spectral lines.  This complex dependence actually is an advantage: Since the exact line displacements depend on so many parameters, stellar atmospheric properties can be constrained by many different wavelength measures for various lines in the same star, or between different stars.

The longer-term outlook appears quite promising since the accuracies for astrometric radial velocities will improve with planned future space astrometry missions (e.g., the ESA project {\it GAIA}).  For stars in nearby clusters, the improvement will perhaps not be very dramatic (the accuracy then being limited by cluster kinematics), but more distant clusters will become accessible, offering samples of more unusual (but astrophysically important) spectral types.

For a further discussion, see links from {\rm http://www.astro.lu.se/\~{}dainis/}

\acknowledgments

This work is supported by The Swedish Research Council, the Swedish National Space Board, and The Royal Physiographic Society in Lund.

\end{document}